\documentclass[doublespacing]{elsart}

\usepackage[dvips]{graphicx}

\usepackage{amssymb}
\usepackage{amsmath}
\usepackage{bm}

\begin{document}

\begin{frontmatter}



\title{Fast switching NMR system for measurements of ground-state quadrupole moments of short-lived nuclei}

\author[label1]{K. Minamisono},
\author[label1,label2]{R. R. Weerasiri\thanksref{label5}},
\author[label1,label2]{H. L. Crawford},
\author[label1,label2]{P. F. Mantica},
\author[label3]{K. Matsuta},
\author[label4]{T. Minamisono},
\author[label1,label2]{J. S. Pinter} and
\author[label1,label2]{J. B. Stoker}
\address[label1]{National Superconducting Cyclotron Laboratory, Michigan State University, East Lansing, MI 48824, USA}
\address[label2]{Department of Chemistry, Michigan State University, East Lansing, MI 48824, USA}
\address[label3]{Department of Physics, Osaka University, Toyonaka, Osaka 560-0043, Japan}
\address[label4]{Department for the Application of Nuclear Technology, Fukui University of Technology, Gakuen, Fukui 910-8505, Japan}
\thanks[label5]{Current address: Technova Corp. Lansing, MI 48906, USA}


\begin{abstract}
A $\beta$-ray detecting nuclear quadrupole resonance system has been developed at NSCL/MSU to measure ground-state electric quadrupole moments of short-lived nuclei produced as fast rare isotope beams.  This system enables quick and sequential application of multiple transition frequencies over a wide range.  Fast switching between variable capacitors in resonance circuits ensures sufficient power delivery to the coil in the $\beta$-ray detecting nuclear magnetic resonance technique.  The fast switching technique enhances detection efficiency of resonance signals and is especially useful when the polarization and/or production rate of the nucleus of interest are small and when the nuclear spin is large.
\end{abstract}

\end{frontmatter}

\section{Introduction}
\label{introduction}

Electromagnetic-nuclear moments are one of the most important fundamental properties of a nucleus and provide key information on nuclear structure.  The electromagnetic moments are one-body operators acting on a single nuclear state and are very sensitive to the wave function of a specific nuclear state.  They are complementary to transition moments, which are two body operators acting on two different nuclear states.  

Recent developments in the production of nuclear spin polarization as well as in the production of radioactive-nuclear beams have made it possible to perform studies on nuclear moments away from the stability line in the nuclear chart.  Fast rare isotope beams are considered here and the polarization is produced in projectile-fragmentation \cite{asahi90} or nucleon pick-up \cite{groh03} reactions.  Nuclear moments of such unstable nuclei, whose production rates become small as the nuclei move further from the stability line, are measured with the $\beta$-ray detecting Nuclear Magnetic Resonance ($\beta$-NMR) technique \cite{minamisono74,asahi84}.  

The $\beta$-NMR technique is a highly precise and sensitive technique and requires at least 10$^3$ spins to generate the NMR signal compared to greater than 10$^{17}$ spins in the conventional NMR technique.  This is because the NMR signal in the $\beta$-NMR technique is obtained by detecting the radioactivity from the nucleus, namely from the asymmetric $\beta$-ray angular distribution from the decaying polarized nuclei.  Therefore, the $\beta$-NMR technique is widely used in a variety of fields, for example, in research on nuclear structure\cite{minamisono06, mertzimekis06}, fundamental symmetries of the weak interaction \cite{minamisono01} and condensed matter systems \cite{matsuta99}. 

Measurements of magnetic-dipole moments of unstable nuclei were successfully performed at National Superconducting Cyclotron Laboratory at Michigan State University (NSCL/MSU) to study nuclear structure and charge symmetry of mirror nuclei \cite{minamisono06, mertzimekis06}.  As a next step, electric quadrupole moment measurements are planned to determine nuclear charge distributions and provide information on the deformation of exotic nuclei.  The challenge in the measurements of quadrupole moments is the small NMR signal scattered over a wide range of radio frequencies (RF) due to an electric-quadrupole interaction between the quadrupole moment of nucleus and the electric field gradient.  A strong rotating magnetic field ($H_1$) needs to be applied for an efficient $\beta$-NMR measurement.  The $H_1$ must be maintained over a wide range of transition frequencies and a short period of time.  The latter requirement is established by the short-lived nature of the exotic nucleus of interest.  

A $\beta$-ray detecting Nuclear Quadrupole Resonance ($\beta$-NQR) system has been developed at NSCL based on an existing $\beta$-NMR system \cite{mantica97}.  The $\beta$-NQR technique \cite{minamisono93} is a multiple RF NMR technique used to significantly enhance the small NMR signal.  The potential of this technique was demonstrated, for example, in Ref. \cite{minamisono93b, borremans05}.  The system is especially important for nuclei with low production rates, small polarization and/or large nuclear spin.  The present system enables quick and sequential application of RF scattered over a wide range, with the capability to measure magnetic and quadrupole moments of short-lived radioisotopes produced through intermediate-energy reactions.  

The electromagnetic interaction of magnetic and quadrupole moments with external fields is briefly reviewed in section 2.  The principle of the $\beta$-NQR technique is discussed in section 3 and each component of the system is then detailed in section 4.  The result of verification tests of the $\beta$-NQR system at NSCL is discussed in section 5.

\section{Electromagnetic interaction}
\label{electromagneticint}
\begin{figure}
\begin{center}
\includegraphics[width=12cm,keepaspectratio,clip]{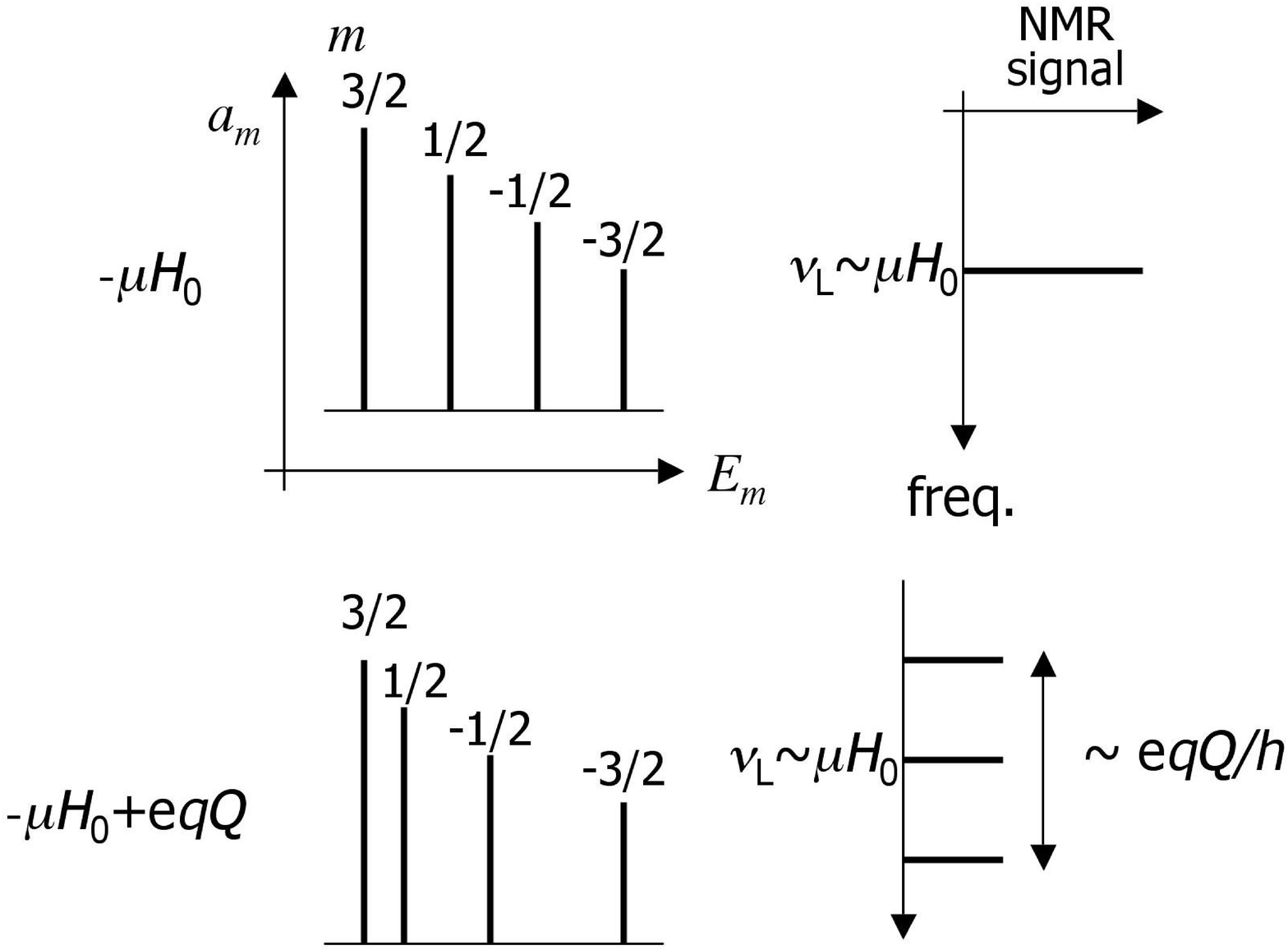}
\end{center}
\caption{Energy levels and transition frequencies for an $I$ = 3/2 nucleus.  In the upper part, energy levels are shown in the presence of a purely magnetic interaction.  A single resonance frequency ($\nu_L$, the Larmor frequency) results due to the evenly-spaced energy levels.  When the electric interaction is added, as shown in the lower part, the spacings between adjacent sublevels become uneven and 2$I$ different resonance frequencies result.}
\label{fig:transitionfreq}
\end{figure}

The hamiltonian of the electromagnetic interaction between nuclear moments and external fields  \cite{abragam86} is given by
\begin{equation}
H = -\bm{\mu}\cdot\bm{H}_0+\frac{eqQ}{4I(2I-1)}\{3I_Z^2-I(I+1)+\frac{\eta}{2}(I_+^2+I_-^2)\}.
\label{eq:hamiltonian}
\end{equation}
Here $\bm\mu$ is the magnetic moment, $\bm{H_0}$ is  the external dipole-magnetic field, $I$ is the nuclear spin, $Q$ is the quadrupole moment, $I_z$ is the third component of the spin operator and $I_{\pm}$ are the raising and lowering operators.  The biggest component of the electric field gradient is defined by $q = V_{ZZ}$ where $V$ is the electrostatic potential and $V_{ii} = d^2V/di^2$.  The asymmetry parameter of the electric field gradient is defined as $\eta = (V_{XX}-V_{YY})/V_{ZZ}$ with $|V_{XX}| < |V_{YY}| < |V_{ZZ}|$.  An appropriate electric field gradient is provided by the internal field of a single crystal.  The energy levels are given to first-order of the electric coupling constant $eqQ/h$ as 
\begin{equation}
E_m = -g\mu_NH_0m + \frac{h\nu_Q}{12}(3\cos^2\theta-1+\eta\sin^2\theta\cos2\phi)\{3m^2-I(I+1)\},
\label{eq:energy}
\end{equation}
where, for simplicity, the electric interaction is regarded as a perturbation to the main magnetic interaction.  In Eq. (\ref{eq:energy}), $m$ is the magnetic quantum number, $\nu_Q = 3eqQ/\{2I(2I-1)h\}$ is a normalized electric coupling constant, and $\theta$ and $\phi$ are the Euler angles between the principal axes of the electric field gradient and the external dipole-magnetic field, respectively.  The first term in Eq. (\ref{eq:energy}) gives the $2I + 1$ magnetic sublevels separated by a fixed energy value determined from the applied dipole-magnetic field and size of the nuclear $g$ factor due to the magnetic interaction (Zeeman splitting).  These sublevels are further shifted by the electric interaction and the energy spacing between adjacent sublevels is no longer constant.  The 2$I$ separate transition frequencies appear as
\begin{equation}
f_{m-1\leftrightarrow m} = \nu_L - \frac{\nu_Q}{4}(3\cos^2\theta-1+\eta\sin^2\theta\cos2\phi)(2m-1),
\label{eq:frequency}
\end{equation}
since the transition frequencies correspond to the energy difference between two adjacent energy levels ($\Delta m = \pm 1$) in Eq. (\ref{eq:energy}).  Here $\nu_L = g\mu_NH_0/h$ is the Larmor frequency.  The variation in the number and position of resonance frequencies between pure magnetic and both magnetic and electric interactions is schematically shown in Fig. \ref{fig:transitionfreq} for the case of $I$ = 3/2.  

When the electric interaction can not be considered as a perturbation to the magnetic interaction, higher order terms of the electric interaction have to be considered.

\section{$\beta$-NQR technique}
\label{bnqrtechnique}
The $\beta$-NQR technique \cite{minamisono93} is an NMR technique used for short-lived $\beta$-decaying nuclei.  An asymmetric $\beta$-ray angular distribution from nuclear-spin polarized nuclei is utilized to obtain an NMR signal.  The $\beta$-NQR measurement starts with the production of nuclear-spin polarized nuclei.  The polarized nuclei are then implanted into a single crystal within a dipole magnet for the nuclei to be exposed to well-defined electromagnetic fields.  The application of $H_1$ and further manipulation of nuclear spins induces a change in the asymmetric $\beta$-ray angular distribution from the polarized nuclei.  The asymmetry change is detected as a function of applied RF to deduce the magnetic and/or quadrupole moment.  Each of these steps is described in detail below for the specific application of $\beta$-NMR measurements with the fast switching NQR system at NSCL.

Projectile fragmentation or nucleon pick-up reactions are used to produce the nuclei of interest in a spin-polarized state.  A few percent nuclear polarization is typically realized from these reactions \cite{asahi90, groh03}.  Generally, alignment and higher-order tensor polarizations may also be produced, depending on the experimental conditions and the nuclear spin.  Since the $\beta$-NMR technique is sensitive to the polarization (and the odd rank of higher-order tensor polarizations), only polarization is discussed in this work for simplicity.  The nuclear polarization $P$ is defined by a linear distribution of populations in energy levels $E_m$ given in Eq. (\ref{eq:energy}) as $P = \sum a_m m/I$, where $a_m$ is the population in state $E_m$.  The polarized nuclei are implanted into a single crystal surrounded by an RF coil under an external dipole-magnetic field, where the electromagnetic interactions are utilized to probe the nuclear moments as discussed in section \ref{electromagneticint}.  

The $\beta$-ray angular distribution from polarized nuclei is given by $W(\varphi) \sim 1+AP\cos\varphi$ where $\varphi$ is the angle between the polarization direction and the direction of the decay $\beta$ particle momentum, $A$ is the asymmetry parameter and $P$ is the initial polarization produced in the nuclear reaction as defined earlier.  The NMR signal is obtained from a double ratio defined as 
\begin{equation}
R = \left[\frac{W(0^\circ)}{W(180^\circ)}\right]_{\rm off}\Bigg/\left[\frac{W(0^\circ)}{W(180^\circ)}\right]_{\rm on},
\label{eq:ratio}
\end{equation} 
where off(on) stands for without(with) the RF applied.  On resonance, [see Eq. (\ref{eq:frequency})], transitions among two adjacent magnetic sublevels are induced and some populations are equalized, resulting in a reduction of initial polarization.  The NMR signal is approximately given from Eq. (\ref{eq:ratio}) by $R \approx 1 + 2A\Delta P$, where $\Delta P$ is a change of polarization between RF on and off.  The polarization change takes the maximum number $\Delta P = P$ if transitions among all sublevels are saturated and all populations are equalized by the applied RF.  This gives the maximum NMR signal as $R \approx 1 + 2AP$.

The magnetic moment is obtained by locating the single resonance frequency corresponding to the Larmor frequency, the first term in Eq. (\ref{eq:frequency}).  The quadrupole moment, on the other hand, is obtained by locating the $2I$ transition frequencies in Eq. (\ref{eq:frequency}).  The conventional $\beta$-NMR technique relies on detecting each of these $2I$ transition frequencies from independent measurements.  A single frequency is applied and the entire range of RF swept.  The NMR signal is significantly smaller than the maximum signal expected from the initial polarization because the application of one transition frequency induces only a partial reduction of initial polarization.  The maximum NMR signal, on the other hand, can be achieved if all $2I$ transition frequencies in Eq. (\ref{eq:frequency}) are applied, resulting in the total destruction of the initial polarization.  This technique is called the $\beta$-NQR technique.  Note that the nucleus of interest is short lived and the RF has to be applied in a period of time shorter than the lifetime of the nucleus.  Assuming a linear distribution of populations in $E_m$ as $a_m-a_{m-1} = \epsilon$ and $\sum_{m=-I}^{I}a_m = 1$, where $\epsilon$ is a nonzero constant, the $\beta$-NQR technique is $(8/9)I^3(I+1)^2(2I+1)^2$ times more efficient than the conventional $\beta$-NMR technique.   For example, a one-day measurement of the quadrupole moment of an $I$ = 3/2 nucleus with the $\beta$-NQR technique would require 300 days of measurement with the conventional $\beta$-NMR technique.  

It is noted that the $\beta$-NQR technique with zero external magnetic field \cite{salman06} may be a powerful technique to enhance the efficiency of quadrupole moment measurements.  At zero magnetic field, the number of resonance transitions is reduced due to the degeneracy between $E_{\pm m}$ states, as seen in Eq. (\ref{eq:energy}).  The structure of the resonance spectrum becomes simpler than that obtained in the $\beta$-NQR technique with and external dipole-magnetic field.  There is only one transition frequency in the special cases ($I$ = 1, $\eta$ = 0 or $I$ = 3/2) for the measurement of quadrupole moment.  However, multiple transition frequencies appear in general cases, where the present $\beta$-NQR system may play an essential role for reasons discussed in this paper.

\section{$\beta$-NQR apparatus}
\label{bnqrapparatus}

\subsection{Concept}
\label{concept}
Frequency modulation (FM) of each transition frequency is used to search a wide range of electric coupling constants for resonances.  The wide FM is important especially during the initial search for resonances.  The observed polarization after the irradiation time of $H_1$ is given by $\overline{P} = P_0 e^{-wt_{\rm RF}}$, where $P_0$ is the initial polarization, $w$ is the transition probability between two magnetic sub-level populations and $t_{\rm RF}$ is the RF time \cite{abragam86}.  Here a long spin-lattice relaxation time is assumed.  The applied width of the FM will determine the RF power, namely the required strength of $H_1$, to efficiently destroy the initial polarization.  The FM is defined by the modulating frequency $\nu_{\rm FM}$ and time $t_{\rm FM}$, needed to scan the applied FM, as ${\rm FM} = \nu(t_{\rm FM}) - \nu_0 = \nu_{\rm FM}t_{\rm FM}$, where the frequency at time $t$ is given by $\nu(t) = \nu_{\rm FM}t + \nu_0$ with the initial frequency $\nu_0$.  The transition probability is then approximated by 
\begin{align}
w &= \int_{0}^{t_{\rm FM}}\frac{\left(\gamma H_1\right)^2}{2} g(\nu)dt \Big/t_{\rm FM}
=  \int_{\nu_0}^{\nu_0+{\rm FM}}\frac{\left(\gamma H_1\right)^2}{2} g(\nu)d\nu \Big/\left(t_{\rm FM}\nu_{\rm FM} \right)\notag\\
&= \int_{\nu_0}^{\nu_0+{\rm FM}}\frac{\left(\gamma H_1\right)^2}{2} g(\nu)d\nu \Big/{\rm FM} 
\sim \frac{(\gamma H_1)^2}{2{\rm FM}}, 
\label{eq:probability}
\end{align}
where $\gamma$ is the gyro-magnetic ratio, $g(\nu)$ is the shape function of the resonance line, generally given by a Gaussian or Lorentzian function.  A wide FM compared to the line width of the resonance is assumed to obtain the final approximation.
A wide frequency modulation requires high $H_1$ amplitude to keep a fixed magnitude of the transition probability.  

\begin{figure}
\begin{center}
\includegraphics[width=12cm,keepaspectratio,clip]{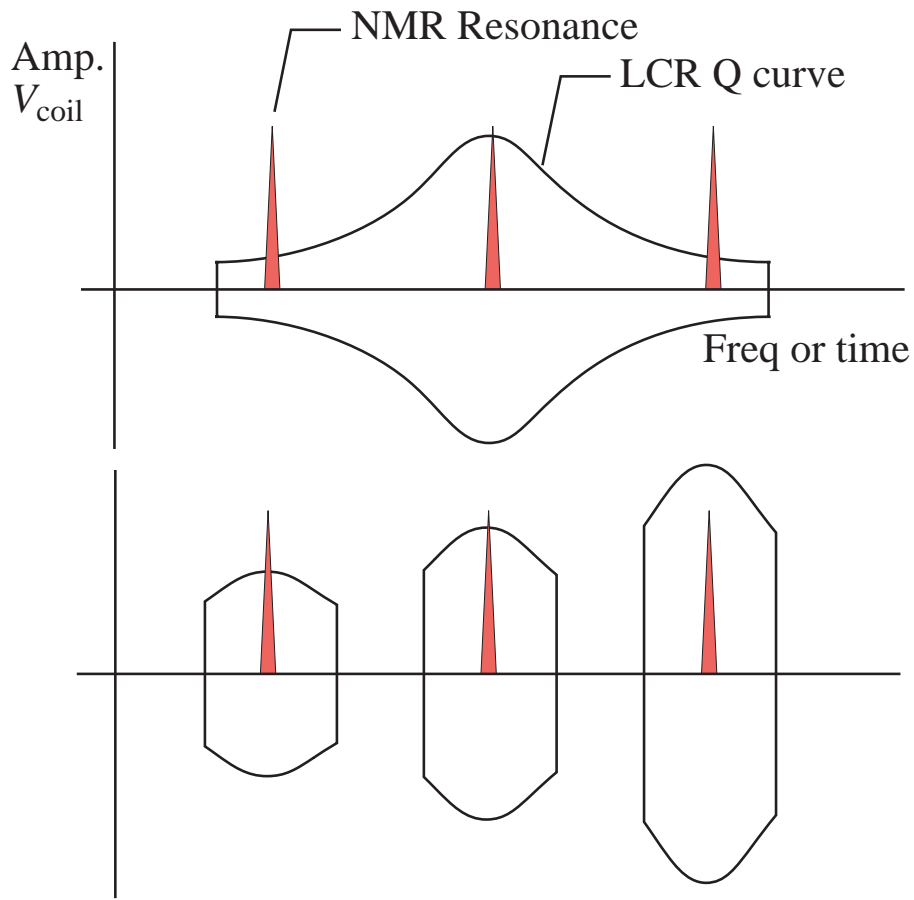}
\end{center}
\caption{Schematic justification of the multiple capacitor system.  The voltage across the RF coil is shown for three different applied frequencies over time.  Sufficient $H_1$ amplitudes over the applied transition frequencies are achieved in the multiple capacitor system as shown by the solid curves (Q curve of LCR resonance circuit) in the lower part, which may not be realized in a single capacitor system as shown in the upper part.}
\label{fig:multicapsys}
\end{figure}

The high $H_1$ amplitude is achieved by use of an LCR resonance circuit, where L is the inductance of the RF coil that produces the $H_1$, C is the capacitance and R is the resistance.  The LCR resonance condition for frequency $f$ is given by $f = 1/(2\pi \sqrt{LC})$.  Multiple variable capacitors are used with fixed L and R to tune the resonance circuit and achieve impedance matching to the RF amplifier.  Such operation ensures sufficient $H_1$ for all transition frequencies within an FM scan.  Transition frequencies are sequentially applied to the LCR resonance circuit by selecting one of the variable capacitors using fast relay switches.  The selected capacitor is tuned to the specific capacitance that satisfies the LCR resonance condition for a particular frequency.  The voltage across the RF coil obtained with the present system is schematically shown in Fig. \ref{fig:multicapsys} as a function of applied frequency.  The locations of NMR resonances for $I$ = 3/2 nuclei are also shown.  Sufficient $H_1$ amplitudes are achieved for three transition frequencies as shown by solid curves in the lower part of the figure.  The solid curves are the Q curves of the LCR resonance circuit and are obtained when an appropriate capacitor is selected by one of the switches for each frequency.  Note that $H_1$ is proportional to the voltage across the coil and inversely proportional to the frequency as $H_1 \propto V/f$.  Sufficient $H_1$ amplitude may not be achieved in a single capacitor system because the applied voltage to the first and third frequencies, which are located at the edge of the Q curve, could be very small as shown in the upper part of Fig. \ref{fig:multicapsys}.  

\begin{figure}
\begin{center}
\includegraphics[width=12cm,keepaspectratio,clip]{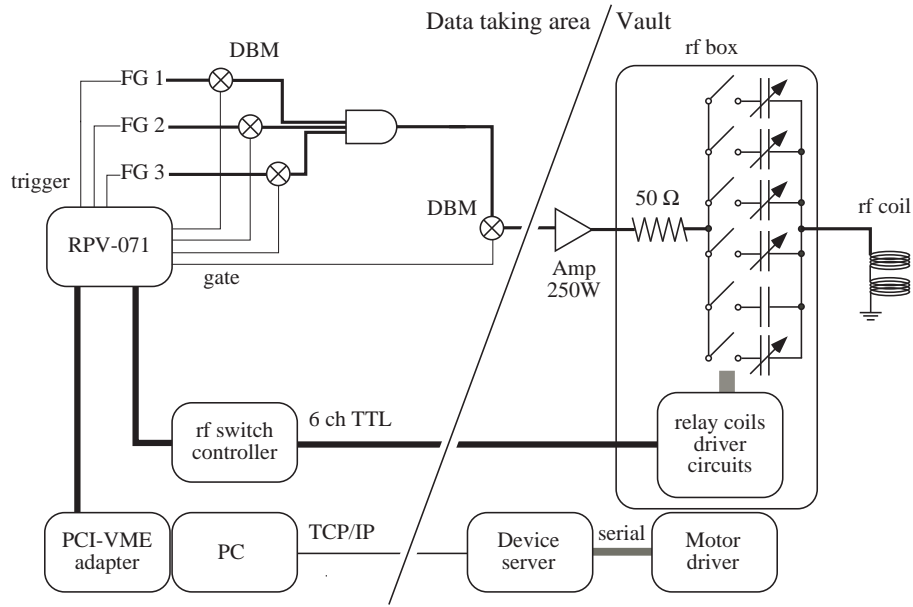}
\end{center}
\caption{Schematic illustration of the $\beta$-NQR system.  One of the RF signals generated by the function generators is selected by a gate and sent to the amplifier.  The amplified signal is then applied to an RF coil through an LCR resonance circuit.  One of six variable capacitors is selected by a fast relay switch in the RF unit to satisfy the resonance condition.  In the schematic, one of the variable capacitors is replaced by fixed capacitor banks.  A pulse pattern generator, RPV-071, controls the function generators, the timing of DBM gating, the fast relay switches and the pulsing of the short-lived radioactive beam.}
\label{fig:rfcircuit}
\end{figure}

\begin{figure}
\begin{center}
\includegraphics[width=12cm,keepaspectratio,clip]{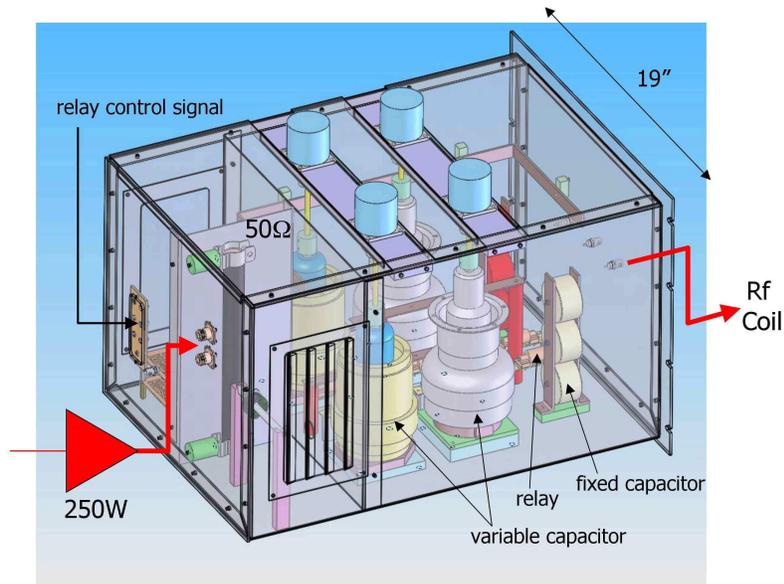}
\end{center}
\caption{Schematic of the $\beta$-NQR RF unit.  The unit contains a 50 $\Omega$ resistor, six capacitors, six relays and control boards for the relay switches.  The unit is connected to the RF coil in a vacuum chamber, completing the LCR resonance circuit.  The variable vacuum capacitors are tuned to a specific capacitance by stepper motors.  Fixed capacitor banks can be included to broaden the range of RF applicability.  The unit is mounted in a standard 19" rack.}
\label{fig:rfunit}
\end{figure}

A schematic of the $\beta$-NQR system at NSCL is shown in Fig. \ref{fig:rfcircuit}.  An RF signal generated by a function generator (labeled FG 1) is selected by a gate (labeled DBM) and sent to the RF amplifier.  The amplified signal is then applied to an RF coil, which is part of the LCR resonance circuit, with an impedance matching resistor or a transformer and one of six variable capacitors.  After some irradiation time (determined from the decay lifetime of the nucleus), the frequency from FG 2 is selected by a gate signal and sent to the same LCR resonance circuit.  A different capacitor, which has been tuned to the second frequency to satisfy the LCR resonance condition, is selected by the fast switching relay system.  Up to six different frequencies can be handled with the current system of 6 function generators, all of which are not shown in Fig. \ref{fig:rfcircuit}.  The components of the resonance circuit are contained in a single unit as pictured in Fig. \ref{fig:rfunit}, with the exception of the RF coil, which is placed in a vacuum chamber surrounding the implantation crystal.  The timing of the DBM, fast switching relays and rare isotope beam pulsing are controlled by a series of trigger signals generated by a VME pulse pattern generator, RPV-071.  Each component of the $\beta$-NQR system is explained in detail in the following subsections.

\subsection{Function generator}
\label{functiongenerator}
RF signals are generated by Agilent Technologies model 33250A digital function generators.  Three function generators are shown in Fig. \ref{fig:rfcircuit} but up to six function generators can be accommodated in the present system, providing access to nuclei with spin up to $I$ = 3.  An RF signal from one of the function generators is selected by a gating signal through a Pulsar Microwave Corp. Double Balanced Mixer model X2M01411B.  The RF signal is then amplified by an EMPOWER model BBS0D3FOQ, 250W, 58 dB RF amplifier and sent to the LCR resonance circuit.  Only one transition frequency is present in the LCR resonance circuit at a time to avoid significant power reflection.  Such reflections cause a distortion of the RF signal, especially when the frequencies are scattered over a wide range and/or when high power is required.  

\subsection{Impedance matching}
Two options are available to impedance match the LCR circuit to the RF amplifier.  Either a 50 $\Omega$ resistor or an impedance matching transformer can be used, depending on experimental requirements.  A higher Q factor of the resonance circuit is obtained with the transformer.  The transformer, therefore, offers higher $H_1$ amplitudes in a narrower frequency region.  A ferrite \#77 manufactured by Amidon is used for the core of the transformer.  The 50 $\Omega$ output impedance of the amplifier is matched to the several $\Omega$ impedance of the resonance circuit at the resonance frequency by adjusting the primary- and secondary-turn numbers of the transformer.  The impedance and turn numbers of the transformer are related as $Z_{\rm s}/Z_{\rm p} = N_{\rm s}^2/N_{\rm p}^2$, where $Z$ and $N$ are the impedance and the turn number of primary (p) and secondary (s) wires, respectively.

\subsection{Capacitor}
Variable vacuum capacitors provide broad tuning capability to different frequencies and have a high power tolerance.  Six capacitors can be implemented in the system and Jennings models UCSXF1500, CVDD1000 and CMV1-4000 vacuum capacitors are used.  The system has 2 of each of the three capacitors with maximum capacitances of 1500, 1000 and 4000 pF, respectively.  Each capacitor can be independently selected by a relay switch to make the LCR circuit resonant with fixed L and R for a specific RF.  Remotely-controlled stepper motors are used to tune the capacitors.  Some of the variable capacitors can be replaced by fixed capacitor banks as shown in Fig. \ref{fig:rfcircuit} when large capacitance ($>$ 10000 pF) is required at lower RF.  The minimum capacitance of the system is determined by a capacitance of the RF unit ($\sim$ 270 pF), which restricts the highest tunable frequency of the system depending on $L$ of the RF coil.  Currently, the system can be tuned to frequencies from 450 kHz to 3 MHz with the $L$ = 10.4 $\mu$H coil.

\subsection{Relay switch}
\begin{figure}
\begin{center}
\includegraphics[width=12cm,keepaspectratio,clip]{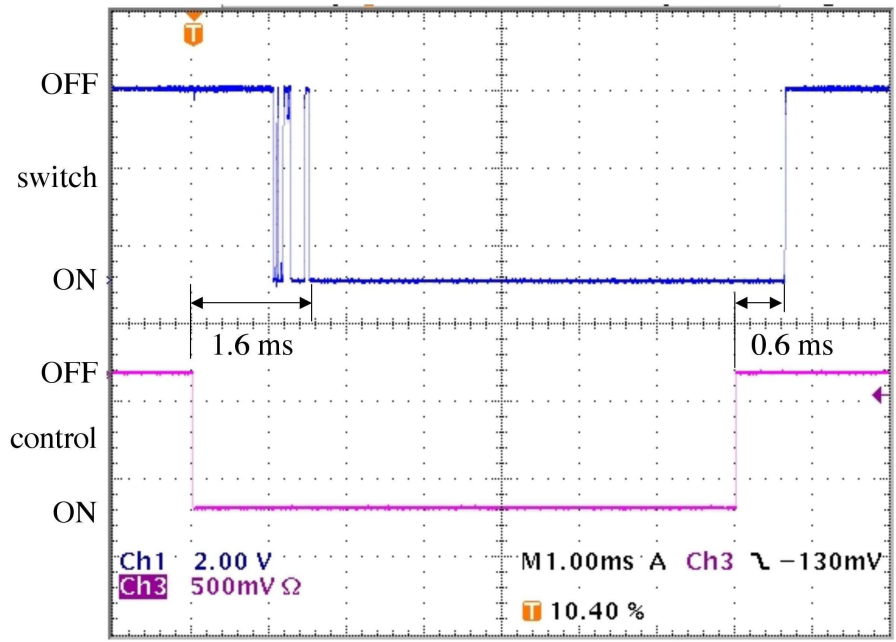}
\end{center}
\caption{Response time of the GIGAVAC model GR6HBA318 fast relay switch.  A DC voltage was applied to the relay and the voltage drop was measured relative to the control signal of the relay coil.  Mechanical bouncing of the relay contact can be seen when the relay turns on.}
\label{fig:switchresponse}
\end{figure}

Fast-switching vacuum relays are employed to provide fast switching speed and fulfill high-voltage and current requirements at a several MHz load frequency.  Currently, GIGAVAC model GR6HBA318 switches are used.   RF signals are applied to the circuit only when the switches are completely on (cold switching) to avoid discharge between relay contacts.  The switching time of the relay, defined by a delay time relative to control signal and including a mechanical bouncing time, was measured as illustrated in Fig. \ref{fig:switchresponse}.  A DC voltage was applied to the relay and the switching speed was measured relative to the control signal applied to the relay coil.  The results are 1.6 ms for off to on and 0.6 ms for on to off.  This performance exceeds manufacturer's specifications (2 ms and 1 ms delay times, respectively).  The few millisecond switching times mean that the $\beta$-NQR system can be used to study nuclei with lifetimes as short as 10's of milliseconds.  The specified mechanical lifetime of the relay is 10$^8$ switching cycles.

\subsection{RF coil}
An RF coil makes up a part of LCR resonance circuit and provides an $H_1$ to the nuclei in the NMR technique.  The coil is located external to the RF unit and surrounds the crystal into which short-lived radioactive nuclei are implanted.  The coil generates a linearly oscillating magnetic field, which can be divided into two oppositely-rotating magnetic fields.  One of the rotating magnetic fields is effective in NMR measurements.  The coil is designed for the specific nucleus, considering the range of RF to be scanned and the required capacitances to tune the LCR circuit.  Currently, two coils arranged in a Helmholtz like geometry are used.  The diameter of each coil is 27 mm, the separation of coils is 22 mm, and the turn number is 22 turns (11 turns for each coil).   The inductance of the coil is $L$ = 10.4 $\mu$H and the coil generates an oscillating $H_1$ of 3 Oe/A$_{\rm DC}$.  

\subsection{Timing control}
Timing control within the system is accomplished by the REPIC model RPV-071, 32 channel output, VME pulse pattern generator with 65k/channel data memory.  A bit pattern is loaded into the memory of the RPV-071 through the VME bus.  The pattern is output-synchronized with an external clock signal.  Each output is used to trigger and/or gate devices.  For example, the function generators, DBM gating for RF signal flow control, fast switching relays and rare isotope beam pulsing are all controlled by the RPV-071 outputs.  Pulse-pattern timing programs are needed to satisfy different experimental objectives.  As an example, one of the existing timing programs is used to determine beam polarization based on pulsed-dipole magnetic field application \cite{anthony00}.  Another program, for the determination of nuclear moments, pulses the RF every 60 s with continuous implantation of the beam.  During RF-on time period, a set of multiple RF pulses are sequentially applied and repeated within this one period.  One cycle of the timing program is loaded to data memory and the cycle is repeated.  Currently, a clock frequency of 2 kHz is used and thus the minimum length of the pulse is 1/2 kHz = 0.5 ms and the maximum length of the pulse or one cycle of timing program is (65k data point)/2 kHz = 32.5 s.  A graphical user interface was developed using Tcl/Tk (scripting language/graphical user interface tool kit) \cite{tcltk} to program and control the pulse pattern generator based on the NSCLDAQ VME Tcl extension \cite{fox06}.  

\section{Verification tests}
\begin{figure}
\begin{center}
\includegraphics[width=12cm,keepaspectratio,clip]{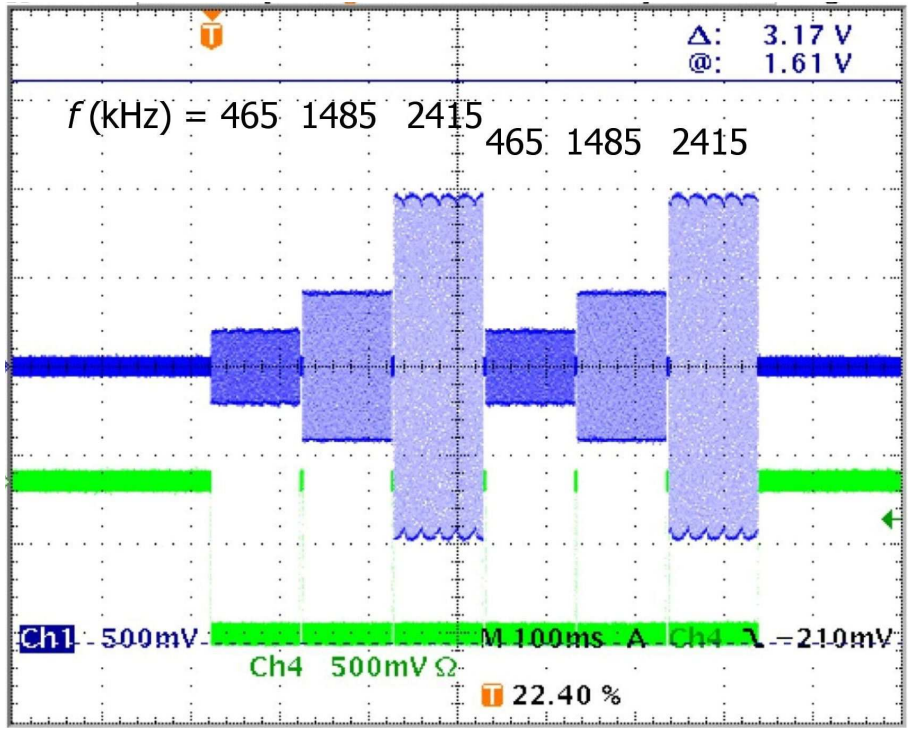}
\end{center}
\caption{Test result of the $\beta$-NQR system.  The partial voltage across the RF coil was monitored by an oscilloscope.  A set of three different frequencies was applied and repeated twice.  Nearly constant $H_1$ among these signals was achieved.  Note that the switching time between signals was 3 ms.}
\label{fig:testresult}
\end{figure}

A test was performed to demonstrate the capability of fast switching and constant $H_1$ amplitudes among multiple transition frequencies.  A set of three independent frequencies was applied sequentially to the $\beta$-NQR system.  The applied signals had frequencies 465, 1485 and 2415 kHz.  These signals were linearly frequency modulated with a modulation width of  50 kHz and a modulation frequency of 50 Hz.  Each signal was maintained for 100 ms and applied in the order of increasing frequency.  The cycle was repeated twice.  A fraction of the voltage across the RF coil was monitored by an oscilloscope as shown in Fig. \ref{fig:testresult}.  Note that the time scale of the oscilloscope is longer than the frequency of signals applied so that only the amplitude profile of the signals is seen.  Constant $H_1$ amplitude was achieved for each of the three signals, since $H_1 \propto V/f$.  The actual voltages across the coil were 360, 1170 and 1920 V$_{\rm pp}$, respectively, and the $H_1$ was deduced to be $\sim$ 8 G.  The 3-ms switching time between two signals did not introduce any discharge and/or distortion to the RF signal during operation.  

\begin{figure}
\begin{center}
\includegraphics[width=12cm,keepaspectratio,clip]{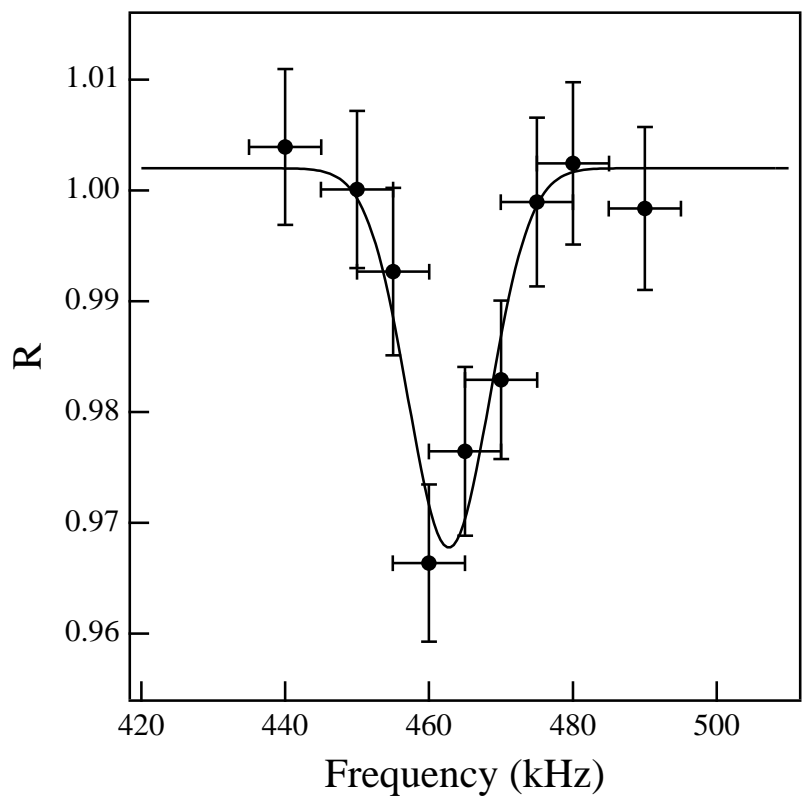}
\end{center}
\caption{The NMR signal for polarized $^{37}$K nuclei implanted in a KBr single crystal.  The solid circles are data and the solid line is a gaussian fit to the data.}
\label{fig:37KinKBr}
\end{figure}

The NSCL $\beta$-NQR system was used in an experiment to measure the well known dipole moment and less precisely known quadrupole moment of $^{37}$K($I^{\pi}$ = 3/2$^+$, $T_{1/2}$ = 1.22 s).  The measurement of the known magnetic moment of $^{37}$K, where no electric interaction is required, was needed to determine the Larmor frequency before the quadrupole moment measurement.  Such a measurement is important since the expected transition frequencies of $^{37}$K in the quadrupole moment measurement are given relative to the Larmor frequency, as shown in Eq. (\ref{eq:frequency}).  The $^{37}$K nuclei were produced through a charge pick-up reaction with a 150 MeV/nucleon $^{36}$Ar primary beam impinging a $^{9}$Be target.  The pick-up reaction was employed because a large nuclear polarization is expected around the central momentum of the $^{37}$K fragment, where the production rate is highest \cite{groh03, groh07}.  The polarized $^{37}$K ions were first implanted into a cubic KBr single-crystal placed at the center of the dipole magnet where the $\beta$-NMR technique was applied.  One of the relays in the $\beta$-NQR system was kept closed and the system worked as a single LCR resonance circuit, since only a single transition frequency was searched.  The RF on/off technique was used to detect the resonance.  The applied RF, during the 60 s on time, was linearly frequency modulated with a modulation width of  50 kHz and a modulation frequency of 50 Hz.  The resulting NMR spectrum is shown in Fig. \ref{fig:37KinKBr}.  A Gaussian function was used to fit the observed resonance and the central frequency was deduced to be 463 $\pm$ 1 kHz.  The magnetic moment was deduced as $|\mu$($^{37}{\rm K})|$ = 0.2026 $\pm$ 0.0005 (stat.) $\pm$ 0.0002 (syst.) $\mu_{\rm N}$, consistent with the known value $\mu$($^{37}{\rm K})$ = +0.20321 $\pm$ 0.00006 $\mu_{\rm N}$ \cite{platen71}.  The external dipole-magnetic field used was $H_0$ = 0.4498 T, which was measured by an NMR probe placed at the position of the KBr implantation crystal.

Based on the NMR signal ($\sim$ 3\%) in the $^{37}$K in KBr measurement, the expected NMR signal in the quadrupole moment measurement would be $\sim 0.3\%$ for each resonance frequency due to the quadrupole splitting.  With the given NMR signal, it would take at least 2700 hours to obtain a nuclear quadrupole resonance spectrum with conventional $\beta$-NMR technique, based on the time (9 hours) used to obtain the $^{37}$K in KBr spectrum shown in Fig. \ref{fig:37KinKBr}.  Here it is assumed that the same level of figure of merit is achieved as the $^{37}$K in KBr.  The figure of merit is defined by $(R-1)^2Y$ (obtained signal squared times counting rate).   The use of the $\beta$-NQR system is essential to perform a reliable and efficient measurement with the full NMR signal.

\begin{figure}
\begin{center}
\includegraphics[width=12cm,keepaspectratio,clip]{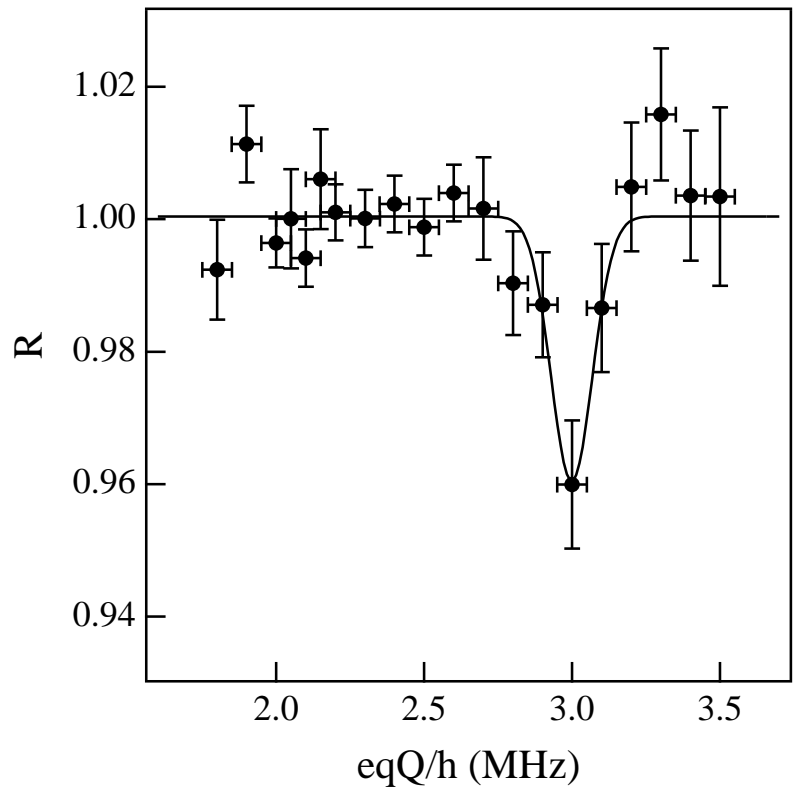}
\end{center}
\caption{Quadrupole resonance spectrum of $^{37}$K in KDP crystal.  The solid circles are data and the solid line is a gaussian fit to the data.}
\label{fig:37KinKDP}
\end{figure}

For the $\beta$-NQR measurement, polarized $^{37}$K ions were implanted into a tetragonal KH$_2$PO$_4$ (KDP) single-crystal.  The $^{37}$K ions are supposed to substitute the K in KDP crystal, where there is an electric field gradient parallel to the c-axis with $\eta$ = 0.  The c-axis was set parallel to the external magnetic field ($\theta = 0$).  The $\beta$-NQR technique was applied and the electric coupling constant of $^{37}$K in KDP crystal was searched by varying three transition frequencies for a given electric coupling constant.  The transition frequencies were obtained by numerically solving the hamiltonian [Eq. (\ref{eq:hamiltonian})], since the electric interaction can not be considered as a perturbation to the magnetic interaction, $\nu_Q > \nu_L$.  Eq. (\ref{eq:frequency}), however, may be used for the system with $\theta = 0$ and $\eta = 0$, where the first order perturbation calculation gives exact transition frequencies.  These three frequencies were repeatedly applied to the implanted $^{37}$K ions in sequence.  Each signal was linearly frequency modulated with a modulation width of  50 kHz and a modulation frequency of 50 Hz and maintained for 100 ms.  The magnetic field strength of each applied frequency was $\sim$ 6 G.  The $^{37}$K ions were continuously implanted into the KDP crystal and a set of three transition frequencies was pulsed every 30 second.  The obtained nuclear quadrupole resonance spectrum is shown in Fig. \ref{fig:37KinKDP}, where the NMR signal $R$ is plotted as a function of the electric coupling constant.  The solid circles are the experimental data and the x-axis value of each point corresponds to the electric coupling constant, which represents the set of three transition frequencies for $I$ = 3/2.  The horizontal bar of each point is the range of electric coupling constant covered by the frequency modulation of the applied frequencies.  The solid line is a Gaussian fit to the data.  The electric coupling constant of $^{37}$K measured in KDP crystal was determined to be $|eqQ/h| = 2.99 \pm 0.07$ MHz from the centroid of the fit.  The corresponding quadrupole moment, $|Q(^{37}$K)$|$ = 10.6 $\pm$ 0.4 efm$^2$, is deduced from $eqQ/h$ \cite{seliger94} and $Q$ \cite{sundholm93} for $^{39}$K.  It took 32 hours to complete the NQR spectrum, which is significantly shorter than the minimum time (2700 hours) required with the conventional $\beta$-NMR technique.  It is noted that the 32 hours includes the time needed to search for the resonance and thus is longer than the naive prediction (9 hours) given above.  

\section{Summary}
\label{summary}
A fast switching $\beta$-NQR system has been developed at NSCL based on an existing $\beta$-NMR system.  The goal is to extend the nuclear moment program to the measurement of quadrupole moments.  The system consists of an LCR resonance circuit with a fixed RF coil, an impedance matching stage (resistor or transformer) and six variable capacitors.  The capacitors can be independently selected by fast switching relay switches.  Multiple transition frequencies are sequentially applied to the LCR resonance circuit, and one capacitor is selected by the fast relay switches to satisfy LCR resonance condition for the applied frequency.  The $\beta$-NQR system enables sequential application of 2$I$ (up to 6) different transition frequencies scattered over a wide range due to the electric quadrupole interaction.  The RF can be applied over a short period of time (3 ms switching time) and with sufficient amplitude ($\sim$ 8 G for 465, 1485 and 2415 kHz signals) to study short-lived nuclei with small nuclear polarization.  The high efficiency of the $\beta$-NQR system was demonstrated in the measurement of electric quadrupole resonance of $^{37}$K, which would have been impossible with the conventional $\beta$-NMR technique.  

\section*{Acknowledgements}
This work was supported in part by the National Science Foundation, Grant PHY06-06007.  The authors would like to express their thanks to the NSCL electronics and RF group for their help in building the RF system.



\newpage

\end{document}